\documentclass[aps,prl, reprint, superscriptaddress,secnumarabic,amssymb,nobibnotes]{revtex4-2}
\usepackage{graphicx}
\usepackage{amsmath}
\usepackage{subfigure}
\usepackage{verbatim}
\usepackage[english]{babel}
\usepackage[utf8]{inputenc}
\usepackage[colorinlistoftodos]{todonotes}
\usepackage[version=3]{mhchem}
\usepackage{chemfig}
\usepackage{cancel}
\usepackage{hyperref}
\usepackage{tabularx}
\usepackage{color}
\usepackage{harpoon}
\usepackage{upgreek}
\usepackage{epstopdf}
\usepackage{bm}
\usepackage{xcolor}

\usepackage{changes}

\makeatletter
\newsavebox{\@brx}
\newcommand{\llangle}[1][]{\savebox{\@brx}{\(\m@th{#1\langle}\)}%
	\mathopen{\copy\@brx\kern-0.5\wd\@brx\usebox{\@brx}}}
\newcommand{\rrangle}[1][]{\savebox{\@brx}{\(\m@th{#1\rangle}\)}%
	\mathclose{\copy\@brx\kern-0.5\wd\@brx\usebox{\@brx}}}
\makeatother

\begin{document}
	
	\title{Enhanced Nonreciprocal Transport through a Coulomb-blockaded Quantum Dot}
	\author{L. Sheng}
	\email{shengli@nju.edu.cn}
	\affiliation{National Laboratory of Solid State Microstructures and Department of
		Physics, Nanjing University, Nanjing 210093, China}
	\affiliation{Collaborative Innovation Center of Advanced Microstructures, Nanjing
		University, Nanjing 210093, China}

	\author{Z. W. Wang}
	\affiliation{National Laboratory of Solid State Microstructures and Department of
		Physics, Nanjing University, Nanjing 210093, China}
	
%
	\author{H. Geng}
\affiliation{College of Physics, Nanjing University of Aeronautics and Astronautics, Nanjing 211106, China}
\affiliation{Key Laboratory of Aerospace Information Materials and Physics (NUAA), MIIT, Nanjing 211106, China}


	\author{D. Y. Xing}
    \email{dyxing@nju.edu.cn}
	\affiliation{National Laboratory of Solid State Microstructures and Department of
		Physics, Nanjing University, Nanjing 210093, China}
	\affiliation{Collaborative Innovation Center of Advanced Microstructures, Nanjing
		University, Nanjing 210093, China}

	\date{\today}

	\begin{abstract}
We investigate nonreciprocal electron transport in an interacting quantum dot 
via a master-equation approach. When mirror and time-reversal symmetries are simultaneously 
broken by Rashba spin–orbit coupling and a Zeeman field, weak nonreciprocity emerges even without Coulomb 
interaction. Intriguingly,  we find that, as the system enters the wide Coulomb-blockade 
regime, the nonreciprocal factor is strongly enhanced with increasing interaction strength. 
Voltage reversal can redistribute the populations of the two 
singly occupied states, yielding constructive contributions to the nonreciprocal response. This minimal 
model opens a new avenue toward strong-correlation-engineered quantum rectification.

	\end{abstract}
	\maketitle	
	

Diodes, the canonical examples of nonreciprocal 
transport, are fundamental building blocks of electronic and information technologies. 
These conventional rectifiers, including p–n junctions and phonon diodes, rely on static spatial asymmetry for operation. 
Recent investigations of nonreciprocal responses in geometrically symmetric systems seek quantum analogs of such classical diode devices,
whose function relies more on material microstructure and internal asymmetry.
Early theoretical and experimental studies of nonlinear mesoscopic transport revealed 
that Onsager–Casimir reciprocity relations forbid nonreciprocal response within the linear-response regime, 
and such constraints can break down far from equilibrium~\cite{Sanchez2004MagneticField}. While S\'{a}nchez and B\"{u}ttiker traced 
such symmetry breaking to magnetic‑field‑dependent internal screening potentials, Spivak and Zyuzin demonstrated 
that disordered mesoscopic metals can host a nonreciprocal nonlinear current scaling as $V^2 B$~\cite{Spivak2004ElectronInteraction}.
This signal vanishes in the noninteracting limit, and requires broken inversion symmetry.
For chaotic quantum‑dot cavities, random‑matrix‑theory work predicted mesoscopic sample‑to‑sample fluctuations of the flux‑antisymmetric nonreciprocal 
nonlinear conductance, accounting for finite temperature, dephasing and interaction effects~\cite{Polianski2006ChaoticQD}.
These predictions were supported by experimental observations of a bias‑antisymmetric and 
magnetic‑field‑antisymmetric nonreciprocal nonlinear conductance component (a signature obtained by 
differentiating the $V^2 B$ nonlinear current with respect to bias) whose magnitude probes 
electron‑interaction strengths~\cite{ZumBuhl2006AsymmetryQD}.
Complementary to quantum‑dot systems, Luttinger‑liquid‑based theory further predicted similar 
interaction‑driven nonreciprocal nonlinear magnetotransport signatures in chiral single‑wall carbon nanotubes, 
where the nonlinear response depends explicitly on tube handedness~\cite{DeMartino2006ChiralNanotube}.

Nonreciprocal transport phenomena have been reported across diverse quantum‑material platforms.
A review by Tokura and Nagaosa summarizes nonlinear nonreciprocal responses realized 
in noncentrosymmetric crystals~\cite{Tokura2018Nonreciprocal}.
In superconducting and Josephson‑junction systems, giant nonreciprocal current–voltage 
characteristics have been predicted in noncentrosymmetric SNS junctions~\cite{Liu2024GiantNonreciprocity}.
Finite‑frequency linear nonreciprocity appears in
multiterminal Josephson junctions through Andreev bound‑state transitions and
Berry‑phase effects~\cite{Virtanen2024NonreciprocalJosephson}, while structurally asymmetric junctions exhibit
nonreciprocal dc characteristics from side‑dependent charging‑energy differences~\cite{Misaki2021TheoryNonreciprocalJosephson}.
Dissipation engineering has been employed to control multiple‑Andreev‑reflection currents in four‑terminal quantum‑dot–superconductor junctions.
Asymmetric coupling in such structures can produce nonreciprocal electron and Cooper‑pair currents~\cite{Damanet2019DissipationEngineering}.
In quantum‑dot circuits, nonreciprocity can be generated by balancing auxiliary‑dot dissipation against coherent
hopping~\cite{Liu2020DissipationQD}, and a general Green's‑function framework has been developed for non‑Hermitian
systems with nonreciprocal quantum dots as representative examples~\cite{Yan2024TransportNonHermitian}. Hydrodynamic
electron fluids can host sizable quadratic nonreciprocal conductance through velocity‑dependent viscosity~\cite{Kirkinis2025Nonreciprocity}.
In chiral magnets, quantum spin fluctuations produce a Kondo‑like logarithmic temperature correction to the nonlinear
conductivity~\cite{Murata2026KondoNonreciprocal}. Nonreciprocal transport, explored in Weyl semimetals with 
engineered Fermi-arc surface states~\cite{Jia2024PRBWeyl1,Jia2025Weyl2}, can also emerge in the ballistic regime 
from asymmetric band structures, where bias-dependent static Coulomb potentials and charge redistribution serve 
as a key ingredient~\cite{Zou2024PRBBallistic,Zou2025NonreciprocalWSM}.
Linear nonreciprocal stochastic models have been studied as a Gaussian‑level theoretical framework, exhibiting enlarged stability
regions and enhanced fluctuations at critical exceptional points~\cite{Shmakov2025GaussianNonreciprocal}.
Reservoir‑engineered nonreciprocity has further been explored to enhance energy storage in quantum batteries~\cite{Ahmadi2024NonreciprocalBatteries}.

Nonlinear transport under finite bias lies beyond the simple single-particle picture, even for few-level systems without 
interactions, where shared reservoir coupling leads to nontrivial dynamic population redistribution among population configurations.
While previous works have revealed multiple mechanisms for nonreciprocal transport, the impact of many-body statistical 
correlations on asymmetric conduction in geometrically symmetric nanostructures, particularly in the presence of 
strong interaction, has not been fully elucidated.

In this Letter, we systematically investigate nonreciprocal transport in a quantum-dot system via 
numerically exact master-equation simulations. We demonstrate that the coexistence of Rashba 
spin–orbit coupling and Zeeman field, breaking mirror and time-reversal symmetries, gives rise to 
weak intrinsic nonreciprocity even without Coulomb interaction. Remarkably, strong many-body 
correlations emerging in the broad Coulomb-blockade regime~\cite{Beenakker1991,Kouwenhoven1997,Aleiner2002} 
significantly amplify the transport asymmetry. We show that bias-modulated population exchange 
between spin-resolved singly occupied states, without net charge accumulation, serves as the 
dominant origin of the nonreciprocal response.
This minimal quantum-dot platform lays a foundation for both theoretical understanding and potential 
practical applications of the strong-correlation-driven nonreciprocal quantum transport.

\begin{figure}[htbp]
\centering
\includegraphics[width=0.74\columnwidth]
{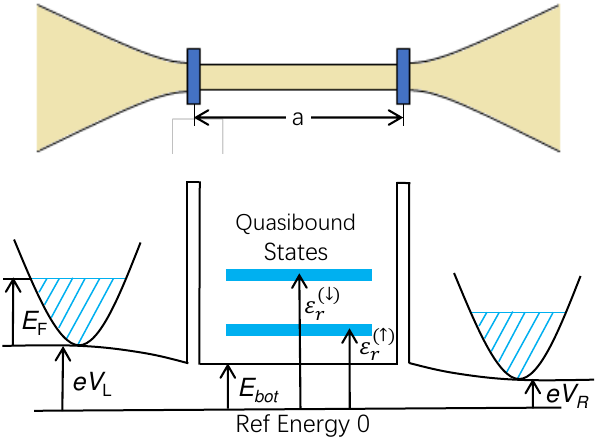}
\caption{(Color online) Schematic of a quantum dot formed by an 
adiabatically constricted quasi-one-dimensional channel. The dot is 
coupled to two reservoirs via $\delta$-function tunnel barriers of 
height $V_{\text{barrier}}$, with independent bias voltages $V_L$ and 
$V_R$ applied to the left and right leads.}
\label{fig:fig1}
\end{figure}

Let us consider a quantum dot formed by an 
adiabatically constricted quasi-one-dimensional channel, as shown in Fig.\ \ref{fig:fig1}.
In the presence of Rashba spin-orbit coupling and Zeeman field,
the lowest quasibound state near the Fermi level in the quantum dot is split into two energy
levels $\varepsilon_{r}^{(s)}=E_{bot}+E_{0}\mp B$ with spin index $s=\uparrow$ and $\downarrow$.
$E_{bot}$ is the energy of the band bottom in
the cavity, which incorporates the effect of gate voltage and work-function 
difference between the cavity and the electrodes. $E_{0}$ is the energy of the 
resonant state measured from the band bottom in the absence of Zeeman splitting
and Rashba spin–orbit coupling.
$B$ is the strength of the Zeeman field. The Rashba spin-orbit coupling $\alpha_R$
only leads to second-order correction to electron energy and 
can be absorbed into the energy $E_{bot}$, while its primary effect enters the tunneling rates. 
Other quasibound states are assumed to be sufficiently far from the Fermi level and thus negligible for transport.

As derived in Supplemental material from the multiple-reflection dynamics inside the double-barrier cavity,
the spectral function of the spin-resolved quasibound state takes the Lorentzian form
$A^{(s)}(\varepsilon)
=\Gamma_{\rm tot}^{(s)}\big/\pi\big[\big(\varepsilon-\varepsilon_{r}^{(s)}\big)^2+\big(\Gamma_{\rm tot}^{(s)}\big)^2\big]$,
with decay width $\Gamma_{\rm tot}^{(s)}=\Gamma^{(s)}_{L,\text{out}}+\Gamma^{(s)}_{R,\text{out}}$.
$\Gamma^{(s)}_{\gamma,\text{out}}$ denotes the bare tunneling rates
of a spin-$s$ electron in the cavity to reservoir $\gamma$.
The bare tunneling couplings $\Gamma_{\gamma,\text{in}}^{(s)}$, $\Gamma_{\gamma,\text{out}}^{(s)}$ 
are evaluated from standard quantum-mechanical calculations; see Appendix A for details, which are dependent on 
resonant electron energy $E=\varepsilon_{r}^{(s)}$, Rashba spin-orbit
coupling $\alpha_{R}$, Zeeman field $B$, and barrier height $V_{\text{barrier}}$.

A configuration of the quasibound state is labelled by the occupation vector
\[
\boldsymbol{n}=(n_{\uparrow},n_{\downarrow}),
\]
where $n_s$ denotes the occupation of the spin-$s$ quasibound state.
The total energy for a given charge configuration reads
\begin{equation}
E(\boldsymbol{n})
=\sum_{s=\uparrow,\downarrow} n_s \varepsilon_{r}^{(s)}
+U\,n_{\uparrow}n_{\downarrow},
\end{equation}
where $n=n_{\uparrow}+n_{\downarrow}$,
and $U$ describes the on-site Coulomb repulsion inside the cavity.
We define transition energy as
\begin{equation}
\begin{aligned}
\Delta E_{\text{add}}(\boldsymbol{n},s)
&=E(\boldsymbol{n}+\boldsymbol{\delta}_{s})-E(\boldsymbol{n}),\\
\Delta E_{\text{rmv}}(\boldsymbol{n},s)
&=E(\boldsymbol{n})-E(\boldsymbol{n}-\boldsymbol{\delta}_{s}).
\end{aligned}
\end{equation}
The symbol $\boldsymbol{\delta}_{s}$ denotes the unit occupation increment for the spin-$s$ state.
$\boldsymbol{n}+\boldsymbol{\delta}_{s}$ corresponds to adding one electron to the spin-$s$ level,
while $\boldsymbol{n}-\boldsymbol{\delta}_{s}$ removes one electron from this level.

The electron energies are defined relative to the band bottom of the leads 
in the equilibirium state (i.e., $V_{L}=V_{R}=0$), which is chosen as the zero of energy.
When reservoir of lead $\gamma$ ($\gamma=L,R$) is held at electrostatic 
potential $V_\gamma$, all single-particle energy levels and the Fermi level 
of the reservoir shift uniformly by $eV_\gamma$, where $e<0$ denotes the electron charge.
Let $\varepsilon' = \varepsilon + eV_\gamma$ denote the electron energy in the 
global energy reference frame, where $\varepsilon$ is the local energy inside reservoir $\gamma$.
The Fermi–Dirac distribution in the global energy reference frame reads
\begin{equation*}
f_\gamma(\varepsilon')
=\frac{1}{e^{\frac{\varepsilon'-(E_F+eV_\gamma)}{k_B T}}+1}
=\frac{1}{e^{\frac{\varepsilon-E_F}{k_B T}}+1}
\equiv f_0(\varepsilon).
\end{equation*}
The uniform potential shift cancels identically in the Fermi–Dirac argument, 
such that the reservoir occupation reduces to the bias-independent equilibrium distribution $f_0(\varepsilon)$.
As a result, the metallic reservoir retains its bulk charge neutrality.
In contrast, semiconductor p–n junctions exhibit substantial charge redistribution and recombination under bias, which are outside the scope of this work.
Energy conservation during tunneling requires matching the global electron energy $\varepsilon'=\varepsilon+eV_\gamma$ to the transition 
energy of the cavity, which is encoded in the spectral function.

The occupation-dependent tunneling rates read
\begin{equation}
\begin{aligned}
R^{(s)}_{\gamma,\text{in}}(\boldsymbol{n})
&=\Gamma^{(s)}_{\gamma,\text{in}}\int_{0}^{\infty} d\varepsilon \,
f_0(\varepsilon) \\
&\times A^{(s)}\big((\varepsilon+eV_\gamma)-\Delta E_{\text{add}}(\boldsymbol{n},s)\big),\\[4pt]
R_{\gamma,\text{out}}^{(s)}(\boldsymbol{n})
&=\Gamma^{(s)}_{\gamma,\text{out}}\int_{0}^{\infty} d\varepsilon \,
\big[1-f_0(\varepsilon)\big] \\
&\times A^{(s)}\big((\varepsilon+eV_\gamma)-\Delta E_{\text{rmv}}(\boldsymbol{n},s)\big).
\end{aligned}
\end{equation}
This integral describes the convolution between the lead occupation statistics and the cavity resonance spectral density.

The master equation for the configuration probability $P(\boldsymbol{n},t)$ is 
\begin{equation}
\begin{split}
\partial_t P(\boldsymbol{n},t)&=
\sum_{\gamma=L,R}\sum_{s=\uparrow,\downarrow}
\Big[
R^{(s)}_{\gamma,\mathrm{in}}(\boldsymbol{n}-\boldsymbol{\delta}_s)P(\boldsymbol{n}-\boldsymbol{\delta}_s,t)\\
&+R^{(s)}_{\gamma,\mathrm{out}}(\boldsymbol{n}+\boldsymbol{\delta}_s)P(\boldsymbol{n}+\boldsymbol{\delta}_s,t) \\
&-R^{(s)}_{\gamma,\mathrm{in}}(\boldsymbol{n})P(\boldsymbol{n},t)
-R^{(s)}_{\gamma,\mathrm{out}}(\boldsymbol{n})P(\boldsymbol{n},t)
\Big].
\end{split}
\label{eq:4}
\end{equation}
In the steady state, the probability distribution satisfies $\partial_t P(\boldsymbol{n},t) = 0$.
The resulting linear master-equation system is coupled and generally unsolvable via analytical methods.
Therefore, we solve the master equation via standard linear-algebra routines numerically to obtain the stationary 
probability distribution $P(\boldsymbol{n})$.

\begin{figure}[htbp]
\centering
\includegraphics[width=0.96\columnwidth]
{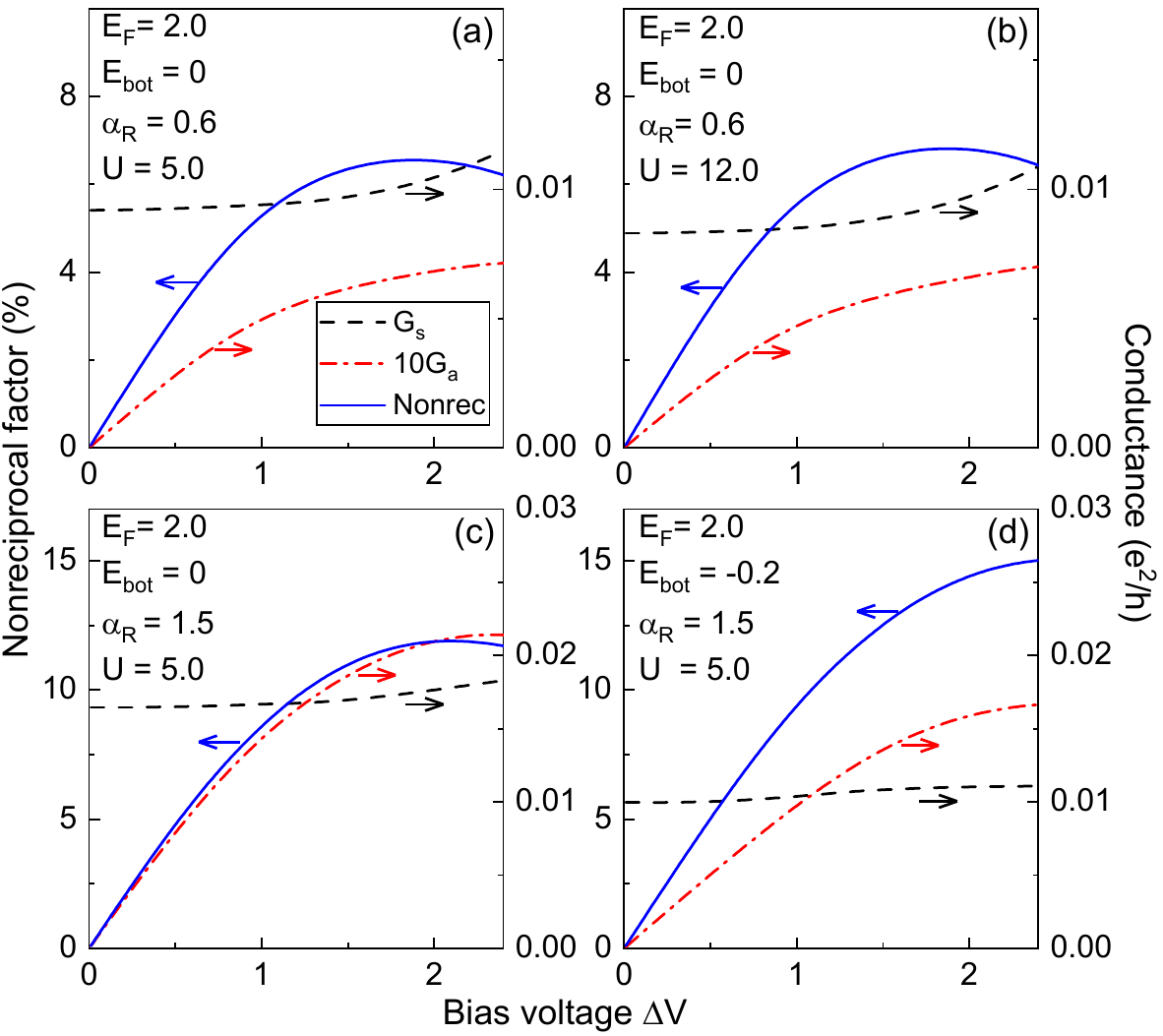}
\caption{(Color online) Nonreciprocal factor, symmetric
conductance $G_{s}$ and antisymmetric conductance $10G_{a}$ versus 
bias voltage $\Delta V$ for several parameter sets. 
Fixed parameters are $V_{\text{barrier}}=5$, $B=0.5$, and $E_{0}=1$.}
\label{fig:fig2}
\end{figure}

With the stationary occupation probabilities determined, we define the total charge current flowing from reservoir $\gamma$ into the cavity as
\begin{equation}
\begin{aligned}
I_\gamma &= e\sum_{s=\uparrow,\downarrow}\sum_{\boldsymbol{n}}
\Big[
R_{\gamma,\text{in}}^{(s)}(\boldsymbol{n}-\boldsymbol{\delta}_s)\,P(\boldsymbol{n}-\boldsymbol{\delta}_s)
\\
&-R_{\gamma,\text{out}}^{(s)}(\boldsymbol{n}+\boldsymbol{\delta}_s)\,P(\boldsymbol{n}+\boldsymbol{\delta}_s)
\Big].
\end{aligned}
\end{equation}
Steady-state charge conservation guarantees $I_L + I_R = 0$, and the net current
flowing from the left electrode to the right electrode can be evaluated by $I=(I_{L}-I_{R})/2$.
For a finite bias voltage $V=V_L-V_R$, we define the symmetric and antisymmetric conductances
as 
\begin{equation}
G_{s}=\frac{I(V)-I(-V)}{2V},\qquad
G_{a}=\frac{I(V)+I(-V)}{2V},
\end{equation}
where $I(V)$ denotes the steady‑state current.
Here $G_s$ is even upon bias reversal and corresponds to the conventional conductance,
while $G_a$ is odd upon bias reversal and represents the nonreciprocal transport response.
The nonreciprocal factor is given by $f_{\text{NR}}=G_a/G_s$.

Figure 2 illustrates the nonreciprocal factor, symmetric conductance $G_s$, 
and scaled antisymmetric conductance $10G_a$ as functions of bias voltage. 
We adopt the unit convention $E_0=a=1$.
Panel (a) serves as the reference case with moderate Rashba coupling $\alpha_R=0.6$ 
and Coulomb interaction $U=5.0$, where $E_F=2.0$ lies within the 
Coulomb-blockade regime. The nonreciprocal factor first rises, saturates, and 
then decreases at larger bias. Comparing panel (b) to (a), increasing $U$ 
to 12.0 only mildly enhances nonreciprocity. By contrast, raising $\alpha_R$ 
to 1.5 in panel (c) substantially strengthens the nonreciprocal response. 
Further lowering $E_\text{bot}$ to $-0.2$ in panel (d) gives an additional boost, 
with the nonreciprocal factor reaching about $15\%$ at $\Delta V\approx 2$.

\begin{figure}[htbp]
\centering
\includegraphics[width=0.92\columnwidth]
{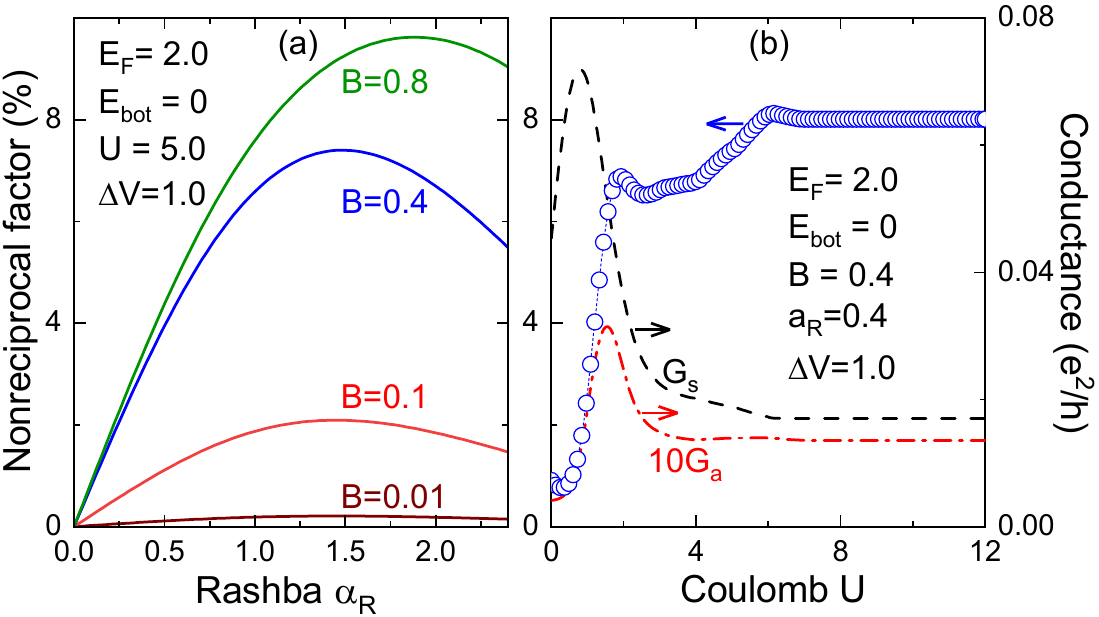}
\caption{(Color online) (a) Nonreciprocal factor as a function of
Rashba coupling $\alpha_R$, and (b) nonreciprocal factor, symmetric
conductance $G_{s}$ and antisymmetric conductance $10G_{a}$ as functions of 
Coulomb interaction $U$. Fixed parameters are 
$V_{\text{barrier}}=5$ and $E_{0}=1$.}
\label{fig:fig3}
\end{figure}
Figure 3(a) presents the nonreciprocal factor as a function of Rashba
spin-orbit coupling for four distinct Zeeman fields B. The nonreciprocal 
factor rises with \(\alpha_R\), reaches a peak, and then declines. 
Larger $B$ yields a stronger nonreciprocal response. Consistent with 
symmetry constraints, the nonreciprocal factor vanishes when \(\alpha_R\to0\) 
or \(B\to0\), as mirror or time-reversal symmetry is restored. Fig.\ 3(b) displays
 the nonreciprocal factor, symmetric
conductance $G_{s}$ and scaled antisymmetric conductance $10G_{a}$ as functions of $U$. 
The nonreciprocal factor increases sharply at small $U$ and 
saturates to a constant value for sufficiently large $U$, where double occupancy 
is fully suppressed and only singly occupied states remain relevant. Minor fine 
structures at intermediate $U$ emerge as double 
occupancy is partially suppressed, arising from different competing tunneling processes
with $U$-dependent transition energies.


\begin{figure}[htbp]
\centering
\includegraphics[width=0.8\columnwidth]
{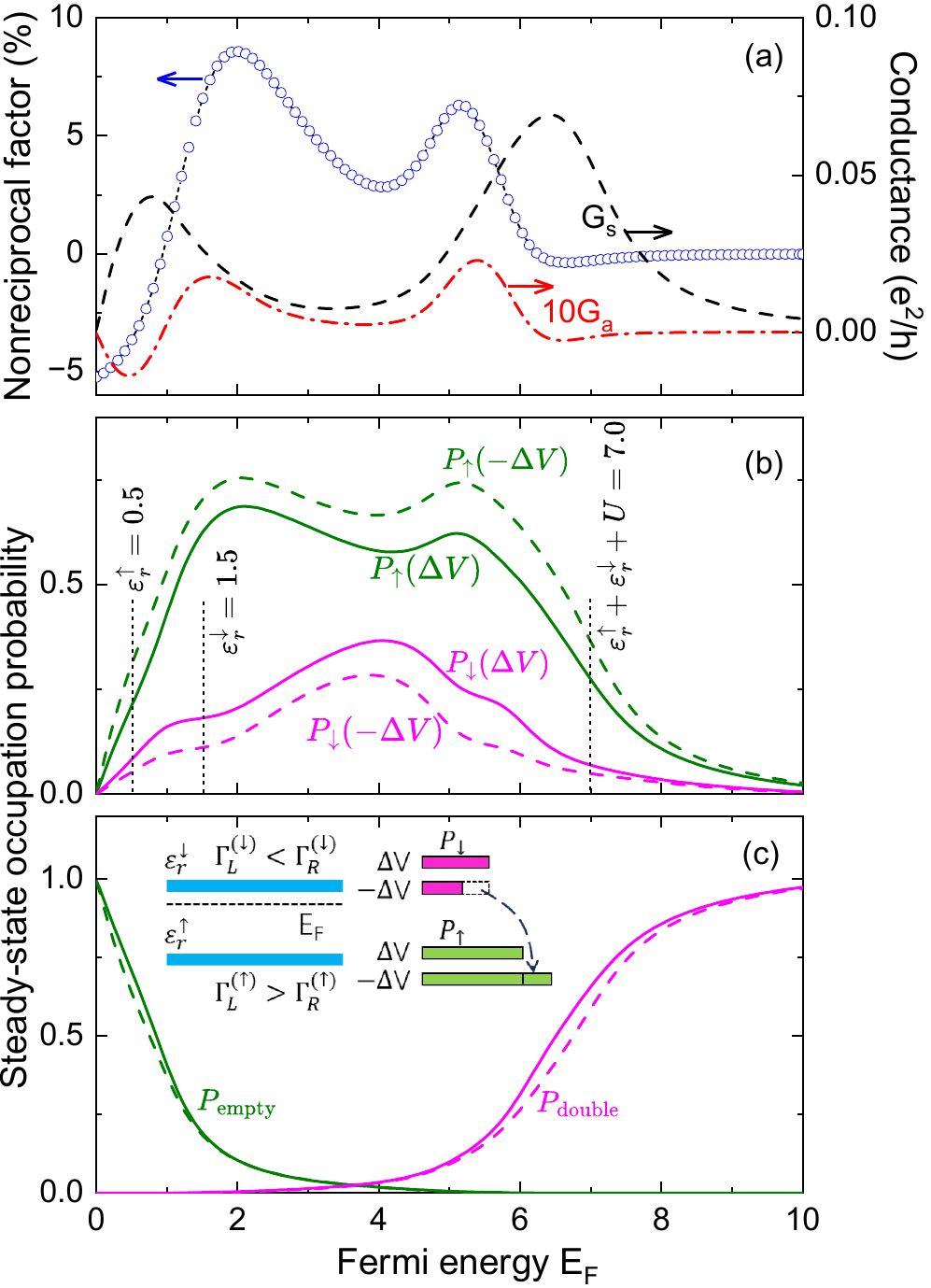}
\caption{(Color online) (a) Nonreciprocal factor, 
symmetric conductance $G_s$, and antisymmetric conductance $10G_a$. (b) Steady-state 
occupation probabilities $P_\uparrow(\pm\Delta V)$ and $P_\downarrow(\pm\Delta V)$ for 
forward and reverse bias. (c) Steady-state probabilities of the empty state $P_{\rm empty}$ and 
doubly occupied state $P_{\rm double}$. Insets show how occupation changes with bias, arising 
from asymmetric left-right tunneling rates.
The parameters are set to $E_{\text{bot}}=0$, $B=0.5$, $\alpha_{R}=1.5$, $U=5$, and $\Delta V=1.0$.
}
\label{fig:fig4}
\end{figure}
Figure 4 illustrates the corresponding transport response and steady-state occupation probabilities 
as we scan the Fermi energy \(E_F\). Panel (a) displays the nonreciprocal factor, symmetric 
conductance \(G_s\), and scaled antisymmetric conductance \(10G_a\) as functions of bias voltage. 
A small negative nonreciprocity emerges as \(U\rightarrow 0\), which arises from symmetry 
breaking combined with finite coupling to the leads (and the associated dephasing; see 
Supplementary Material for further discussion). The nonreciprocal factor becomes positive 
and is strongly enhanced with increasing $U$. Panels (b) and (c) show occupation
probabilities of the many-body configurations: 
$P_\uparrow$ for the $|10\rangle$ configuration (spin-up state singly occupied), 
$P_\downarrow$ for the $|01\rangle$ configuration (spin-down state singly occupied), 
$P_\mathrm{empty}$ for the $|00\rangle$ empty state, and $P_\mathrm{double}$ 
for the $|11\rangle$ doubly occupied state.
$\pm \Delta V$ stands for forward and reverse bias.
The Coulomb-blockade window spans $1.5<E_F<7.0$, with the clean single-occupancy 
regime approximately within $2<E_F<5$. 
For illustration, for this set of parameters, the bare $\Gamma$'s are calculated to be 
$\Gamma^{(\uparrow)}_{L,\text{out}}=\Gamma^{(\uparrow)}_{L,\text{in}}=0.310$, 
$\Gamma^{(\uparrow)}_{R,\text{out}}=\Gamma^{(\uparrow)}_{R,\text{in}}=0.186$, 
$\Gamma^{(\downarrow)}_{L,\text{out}}=\Gamma^{(\downarrow)}_{L,\text{in}}=0.278$, 
$\Gamma^{(\downarrow)}_{R,\text{out}}=\Gamma^{(\downarrow)}_{R,\text{in}}=0.487$.
Microscopic reversibility holds at each interface, $\Gamma^{(s)}_{\gamma,\text{out}}
=\Gamma^{(s)}_{\gamma,\text{in}}$. Rashba spin-orbit coupling breaks mirror symmetry. 
The Zeeman field breaks the time-reversal relation between spin-up left-moving and 
spin-down right-moving states.

Within the Coulomb-blockade regime for $2<E_F<5$, $P_{\text{empty}}$ and $P_{\text{double}}$ are small and 
nearly invariant upon bias reversal, such that $P_\uparrow + P_\downarrow \approx 1$ holds well, 
confining the system to a one-dimensional line segment on the population plane. Motion along this segment 
redistributes probability between the two anticorrelated populations: one rises while the other falls. 
When both \(P_\uparrow\) and \(P_\downarrow\) are appreciable, the two resonant configurations are 
thermodynamically competitive, suggesting that the system’s free energy is flat along this segment. 
Their population ratio becomes susceptible to bias and tunneling asymmetry.
Since the two states carry 
opposite tunneling polarities, their individual nonreciprocal contributions add constructively, 
and a modest redistribution of their weights may yield a large enhancement of the nonlinear 
nonreciprocal response. This sensitive enhancement is absent in the noninteracting counterpart of 
our model: without Coulomb interactions, the system is no longer  
restricted to this one-dimensional 
manifold, additional configurations such as doubly occupied states participate in the dynamics, and the 
clean anticorrelation between these two populations is lost.


To conclude, our results reveal that population competition between singly occupied 
states can constructively amplify nonreciprocal transport under broken symmetries, 
opening a pathway for correlation-dominated asymmetric conductance.
	
	\begin{acknowledgements}
		{\it Acknowledgements}---This work was supported by  the State Key Program for Basic Researches of
		China under Grants No. 2021YFA1400403 (D.Y.X.),
		and the National Natural Science Foundation of China under Grant
		No. 12174182 (D.Y.X.),
		and No. 12304068 (H. G.). 

{\it Data availability}---Code for this study can be found on a public github repository
\url{https://github.com/lishengback/nonreciprocal-qdot-master}.
	\end{acknowledgements}
	
	\bibliographystyle{apsrev4-2}
	\bibliography{references}

\appendix
\section{Appendix A: Minimal Technical Details}

\subsection{Transmission amplitudes and tunneling rates}
The system consists of a quasi-one-dimensional cavity confined by two \(\delta\)-function 
tunnel barriers, hosting Rashba spin–orbit coupling and a Zeeman field. We assume that the 
electrostatic potential step between the quantum dot and electrodes varies sufficiently slowly 
in space, such that an adiabatic quasi-classical treatment is justified. Within this approximation, 
the bias-induced potential offset merely shifts the energy reference for resonant matching
and does not generate extra reflection between the dot and leads. The transmission coefficient does not depend on the 
bias voltage in the classically allowed energy region.

The good quantum number is $s=\pm1$ for the $\sigma_y$ spin subspace. For the present 
geometry, the magnitudes of wavevectors for right-moving ($+$) and left-moving ($-$) spin-$s$ electron waves in the leads satisfy
$k_{\text{ext},+}^{(s)}=k_{\text{ext},-}^{(s)}=k_{\text{ext}}=\sqrt{2mE}/\hbar$.
The group velocity in the leads is $v_{\text{ext}}=\hbar k_{\text{ext}}/m$.
The magnitudes of wavevectors inside the cavity read
\begin{equation}
\begin{aligned}
k_+^{(s)} &= \frac{1}{\hbar}\left(\sqrt{m^2\alpha_R^2 + 2 m s B + 2 m (E-E_{\text{bot}})}-s m\alpha_R\right),\\
k_-^{(s)} &= \frac{1}{\hbar}\left(\sqrt{m^2\alpha_R^2 + 2 m s B + 2 m (E-E_{\text{bot}})}+s m\alpha_R\right).
\end{aligned}
\end{equation}
The group velocities for the cavity modes are
\begin{align}
v_+^{(s)} &= \frac{\hbar}{m}k_+^{(s)} + s\alpha_R,\\
v_-^{(s)} &= \frac{\hbar}{m}k_-^{(s)} - s\alpha_R.
\end{align}
The round-trip traversal time inside the cavity is
\begin{equation}
\tau_{\text{round}}^{(s)} = \frac{a}{v_+^{(s)}} + \frac{a}{v_-^{(s)}}.
\end{equation}

The relevant transmission amplitudes for tunneling rate are
\begin{align}
t_L^{s,\rightarrow} &= \frac{2ik_{\text{ext}}}{i\big(k_{+}^{(s)}+k_{\text{ext}}\big) - \dfrac{2m V_{\text{barrier}}}{\hbar^2}},\\
t_L^{s,\leftarrow} &= \frac{i(k_+^{(s)}+k_-^{(s)})}{i\big(k_+^{(s)}+k_{\text{ext}}\big) - \dfrac{2m V_{\text{barrier}}}{\hbar^2}},\\
t_R^{s,\leftarrow} &= \frac{2ik_{\text{ext}}}{i\big(k_{\text{ext}}+k_{-}^{(s)}\big) - \dfrac{2m V_{\text{barrier}}}{\hbar^2}},\\
t_R^{s,\rightarrow} &= \frac{i(k_+^{(s)}+k_{-}^{(s)})}{i\big(k_{\text{ext}}+k_{-}^{(s)}\big) - \dfrac{2m V_{\text{barrier}}}{\hbar^2}}.
\end{align}
The outward and inward tunneling rates are
\begin{align}
\Gamma_{L,\text{out}}^{(s)} &= \frac{\hbar}{\tau_{\text{round}}^{(s)}}\frac{v_{\text{ext}}}{v_-^{(s)}} \big|t_L^{s,\leftarrow}\big|^2, \\
\Gamma_{R,\text{out}}^{(s)} &= \frac{\hbar}{\tau_{\text{round}}^{(s)}}\frac{v_{\text{ext}}}{v_+^{(s)}} \big|t_R^{s,\rightarrow}\big|^2,\\
\Gamma_{R,\text{in}}^{(s)} &= \frac{\hbar}{\tau_{\text{round}}^{(s)}}\frac{v_-^{(s)}}{v_{\text{ext}}}
\big|t_R^{s,\leftarrow}\big|^2, \\
\Gamma_{L,\text{in}}^{(s)} &= \frac{\hbar}{\tau_{\text{round}}^{(s)}}\frac{v_+^{(s)}}{v_{\text{ext}}} \big|t_L^{s,\rightarrow}\big|^2.
\end{align}

Steady-state transport is described by a master equation governing occupation probabilities of the dot many-body states.
Full transmission amplitudes, tunneling couplings $\Gamma_{\gamma,\text{in}}^{(s)}$
and $\Gamma_{\gamma,\text{out}}^{(s)}$, the general scattering formalism for 
$k_{\text{ext},+}^{(s)}\neq k_{\text{ext},-}^{(s)}$, and the noninteracting 
control calculation are provided in the Supplemental Material.

\end{document}


\title{Supplemental Material\\
Enhanced nonreciprocal transport through a Coulomb blocked quantum dot}

	\author{L. Sheng}
	\email{shengli@nju.edu.cn}
	\affiliation{National Laboratory of Solid State Microstructures and Department of
		Physics, Nanjing University, Nanjing 210093, China}
	\affiliation{Collaborative Innovation Center of Advanced Microstructures, Nanjing
		University, Nanjing 210093, China}

	\author{Z. W. Wang}
	\affiliation{National Laboratory of Solid State Microstructures and Department of
		Physics, Nanjing University, Nanjing 210093, China}
	
%
	\author{H. Geng}
\affiliation{College of Physics, Nanjing University of Aeronautics and Astronautics, Nanjing 211106, China}
\affiliation{Key Laboratory of Aerospace Information Materials and Physics (NUAA), MIIT, Nanjing 211106, China}


	\author{D. Y. Xing}
    \email{dyxing@nju.edu.cn}
	\affiliation{National Laboratory of Solid State Microstructures and Department of
		Physics, Nanjing University, Nanjing 210093, China}
	\affiliation{Collaborative Innovation Center of Advanced Microstructures, Nanjing
		University, Nanjing 210093, China}

\date{}
\maketitle

\renewcommand{\thefigure}{S\arabic{figure}}
\renewcommand{\theequation}{S\arabic{equation}}

\section{Overview}
This supplementary material presents additional derivations, details of the 
transmission‑matrix formalism, control‑case numerical calculations, and supplementary parameter 
scans not contained in the main text and its appendix. 

\section{Minimal model: double‑barrier resonant cavity with Rashba SOC and Zeeman field}
The total Hamiltonian for the one‑dimensional heterostructure reads
\begin{equation}
H = H_0 + H_\text{barrier},
\end{equation}
where the kinetic and spin‑orbit contributions are defined piecewise across the device:
\begin{equation}
H_0=
\begin{cases}
\displaystyle \frac{p_x^2}{2m}\,I, & x<0,\\[4pt]
\displaystyle E_{\text{bot}}+\frac{p_x^2}{2m}\,I + \alpha_R\sigma_y p_x - B\sigma_y, & 0<x<a,\\[4pt]
\displaystyle \frac{p_x^2}{2m}\,I, &  x>a.
\end{cases}
\end{equation}
The barrier term describes two identical delta‑function barriers,
\begin{equation}
H_\text{barrier} = V_{\text{barrier}}\delta(x) + V_{\text{barrier}}\delta(x-a).
\end{equation}
Here $p_x=-i\hbar\partial_x$, $I$ denotes the $2\times2$ identity matrix, and
\[
\sigma_y=\begin{pmatrix}0 & -i\\i & 0\end{pmatrix}.
\]
The Rashba spin‑orbit coupling and a Zeeman field  is assumed to exist in the cavity. 

In both leads and the central cavity, $\sigma_y$ commutes with $H_0$, so $\sigma_y$ serves as a good quantum number 
with eigenvalues $s=\pm1$. For spin‑$s$ electrons in the leads (Zeeman field absent), the dispersion relation is
\[
E = \frac{\hbar^2 k^2}{2m}.
\]
Solving for wavevector at fixed energy yields two roots $k=\pm k_{\text{ext}}$ with $k_{\text{ext}}
=\sqrt{2mE}/\hbar$. To keep the formalism general, we denote $k_{\text{ext},+}^{(s)}$ and $k_{\text{ext},-}^{(s)}$ as 
the magnitude of wavevectors for right‑moving and left‑moving spin‑$s$ electrons in the leads. The physical wavevectors are $+k_{\text{ext},+}^{(s)}$ (right‑going) 
and $-k_{\text{ext},-}^{(s)}$ (left‑going). In our specific model $k_{\text{ext},+}^{(s)}=k_{\text{ext},-}^{(s)}=k_{\text{ext}}$, 
yet the derived formulas remain valid for generalized situations where these wavevector magnitudes differ.

Inside the cavity subject to the Rashba coupling and Zeeman field, the dispersion becomes
\[
E = E_{\text{bot}}+\frac{\hbar^2 k^2}{2m} + s\alpha_R \hbar k - sB.
\]
Solving for $k$ at given energy produces two propagating modes with magnitudes
\begin{equation}
\begin{aligned}
k_+^{(s)} &= \frac{1}{\hbar}\left(\sqrt{m^2\alpha_R^2 + 2 m s B + 2 m (E-E_{\text{bot}})}-s m\alpha_R\right),\\
k_-^{(s)} &= \frac{1}{\hbar}\left(\sqrt{m^2\alpha_R^2 + 2 m s B + 2 m (E-E_{\text{bot}})}+s m\alpha_R\right).
\end{aligned}
\end{equation}
Following the same sign convention, $k_+^{(s)}$ corresponds to right‑moving waves and $k_-^{(s)}$ to left‑moving waves inside the cavity, with physical wavevectors $+k_+^{(s)}$ and $-k_-^{(s)}$.

Continuity conditions at a delta barrier for constant Rashba coupling read
\[
\psi(x^+)=\psi(x^-),\qquad
\psi'(x^+)-\psi'(x^-)=\frac{2m V}{\hbar^2}\psi(x).
\]
Since $\alpha_R$ is spatially constant, the Rashba term integrates to zero across the infinitesimal barrier width. Matching wavefunctions at the left barrier $x=0$ yields the set of reflection and transmission amplitudes
\begin{align}
r_L^{s,\leftarrow} &= \frac{i\big(k_{\text{ext},+}^{(s)}-k_+^{(s)}\big) + \dfrac{2m V_{\text{barrier}}}{\hbar^2}}{i\big(k_{+}^{(s)}+k_{\text{ext},-}^{(s)}\big) - \dfrac{2m V_{\text{barrier}}}{\hbar^2}},\\
t_L^{s,\rightarrow} &= \frac{i(k_{\text{ext},+}^{(s)}+k_{\text{ext},-}^{(s)})}{i\big(k_{+}^{(s)}+k_{\text{ext},-}^{(s)}\big) - \dfrac{2m V_{\text{barrier}}}{\hbar^2}},\\
r_L^{s,\rightarrow} &= \frac{i\big(k_-^{(s)}-k_{\text{ext},-}^{(s)}\big) + \dfrac{2m V_{\text{barrier}}}{\hbar^2}}{i\big(k_+^{(s)L}+k_{\text{ext},-}^{(s)}\big) - \dfrac{2m V_{\text{barrier}}}{\hbar^2}},\\
t_L^{s,\leftarrow} &= \frac{i(k_+^{(s)}+k_-^{(s)})}{i\big(k_+^{(s)}+k_{\text{ext},-}^{(s)}\big) - \dfrac{2m V_{\text{barrier}}}{\hbar^2}}.
\end{align}
Analogous amplitudes for the right barrier located at $x=a$ are
\begin{align}
r_R^{s,\rightarrow} &= \frac{i\big(k_{\text{ext},-}^{(s)}-k_-^{(s)}\big) + \dfrac{2m V_{\text{barrier}}}{\hbar^2}}{i\big(k_{\text{ext},+}^{(s)}+k_{-}^{(s)}\big) - \dfrac{2m V_{\text{barrier}}}{\hbar^2}},\\
t_R^{s,\leftarrow} &= \frac{i(k_{\text{ext},+}^{(s)}+k_{\text{ext},-}^{(s)})}{i\big(k_{\text{ext},+}^{(s)}+k_{-}^{(s)}\big) - \dfrac{2m V_{\text{barrier}}}{\hbar^2}},\\
r_R^{s,\leftarrow} &= \frac{i\big(k_+^{(s)}-k_{\text{ext},+}^{(s)}\big) + \dfrac{2m V_{\text{barrier}}}{\hbar^2}}{i\big(k_{\text{ext},+}^{(s)}+k_{-}^{(s)}\big) - \dfrac{2m V_{\text{barrier}}}{\hbar^2}},\\
t_R^{s,\rightarrow} &= \frac{i(k_+^{(s)}+k_-^{(s)})}{i\big(k_{\text{ext},+}^{(s)}+k_{-}^{(s)}\big) - \dfrac{2m V_{\text{barrier}}}{\hbar^2}}.
\end{align}

For $B=0$, time‑reversal symmetry holds. Time reversal reverses propagation direction and flips spin quantum number. In our centrally 
symmetric double‑barrier geometry, time‑reversed scattering channels correspond to physically comparable scattering events: electrons injected 
from either lead enter the same resonant cavity. This feature translates formal time‑reversal constraints into concrete identities for scattering amplitudes. With $B=0$ one has $k_+^{(s)} = k_-^{(-s)}$, which gives
\begin{align}
t_L^{s,\rightarrow} &= t_R^{-s,\leftarrow}, & r_L^{s,\leftarrow} &= r_R^{-s,\rightarrow},\\
t_L^{s,\leftarrow} &= t_R^{-s,\rightarrow}, & r_L^{s,\rightarrow} &= r_R^{-s,\leftarrow}.
\end{align}
These equalities follow directly from the retarded scattering solutions with consistent boundary conditions and do not require complex conjugation.

In the strong‑barrier limit $\frac{2m V}{\hbar^2}\gg k_{\text{ext},\pm}^{(s)},k_\pm^{(s)}$, 
transmission amplitudes become small, consistent with weak tunneling conditions.

Quasi‑bound states are determined by the self‑consistent round‑trip condition inside the cavity. We define the round‑trip propagation factor for spin‑$s$:
\begin{equation}
\mathcal{L}^{(s)}(E) = r_L^{s,\rightarrow}(E)\, r_R^{s,\leftarrow}(E)\, e^{i\left[k_+^{(s)} + k_-^{(s)}\right]a}.
\end{equation}
Quasi‑bound‑state poles satisfy
\begin{equation}
\mathcal{L}^{(s)}(E) = 1,
\end{equation}
which yields complex resonance energies of the form
\begin{equation}
E = \varepsilon_r^{(s)} - \frac{i}{2}\Gamma_\text{tot}^{(s)}.
\end{equation}
Here $\varepsilon_r^{(s)}$ is the resonant level energy and $\Gamma_\text{tot}^{(s)}$ the total decay width for leakage out through the two barriers.

Within the high‑barrier approximation, reflection amplitudes depend weakly on $\alpha_R$ and $B$, and their phases can 
be approximated by zero‑field values when extracting resonance positions. The total accumulated phase for one full round trip inside the cavity reads
\begin{equation}
\begin{aligned}
\Phi_k &= a\big(k_+^{(s)}+k_-^{(s)}\big)\\
&=\frac{2a}{\hbar}\sqrt{m^2\alpha_R^2+2msB+2m(E-E_{\text{bot}})}.
\end{aligned}
\end{equation}
Upon reflection, electrons reverse propagation direction, so velocity‑linear Rashba contributions cancel. As a consequence, Rashba coupling produces no linear shift of $\varepsilon_r^{(s)}$. Expanding for weak external fields, the Zeeman term survives and gives
\begin{equation}
\varepsilon_r^{(s)} = E_{\text{bot}} + E_{0} - sB + \frac{m\alpha_R^{2}}{2},
\end{equation}
where $E_{0}$ denotes the bare resonance energy relative to the cavity band bottom for $\alpha_R=B=0$. The spin splitting between quasi‑bound levels is approximately $2B$. The Rashba correction enters only at second order and may be absorbed into $E_{\text{bot}}$. Therefore, to linear order in external parameters, the resonance energy shift originates purely from the Zeeman field, while Rashba spin‑orbit coupling modifies transport via wavevectors, group velocities, and tunneling amplitudes. The bare energy $E_0$ acts as a gate‑tunable reference; transport depends on relative energies rather than absolute values.

Group velocities for propagating modes in leads and cavity are summarized below.
For lead modes:
\begin{equation}
\begin{aligned}
v_{\text{ext},+}^{(s)} &= \left.\frac{1}{\hbar}\frac{\partial E}{\partial k}\right\vert_{k=+k_{\text{ext},+}^{(s)}},\\
v_{\text{ext},-}^{(s)} &=-\left.\frac{1}{\hbar}\frac{\partial E}{\partial k}\right\vert_{k=-k_{\text{ext},-}^{(s)}}.
\end{aligned}
\end{equation}
For cavity modes:
\begin{equation}
\begin{aligned}
v_+^{(s)} &=\left.\frac{1}{\hbar}\frac{dE}{dk}\right\vert_{k=k_+^{(s)}}= \frac{\hbar}{m}k_+^{(s)} + s\alpha_R,\\
v_-^{(s)} &=-\left.\frac{1}{\hbar}\frac{dE}{dk}\right\vert_{k=-k_-^{(s)}} = \frac{\hbar}{m}k_-^{(s)} - s\alpha_R.
\end{aligned}
\end{equation}

The energy derivative of the round‑trip phase evaluated at resonance gives
\begin{equation}
\left.\frac{d\Phi}{dE}\right|_{\varepsilon_r^{(s)}} = a\left(\frac{dk_+^{(s)}}{dE}+\frac{dk_-^{(s)}}{dE}\right)
= \frac{1}{\hbar}\tau_{\text{round}}^{(s)},
\end{equation}
where $\tau_{\text{round}}^{(s)}={a}/{v_+^{(s)}}+{a}/{v_-^{(s)}}$ is the round‑trip traversal time inside the cavity.

From small‑parameter expansion of the pole condition, the total decay width is
\begin{equation}
\Gamma_\text{tot}^{(s)} = \frac{\hbar}{\tau_{\text{round}}^{(s)}}\left( \frac{v_{\text{ext},+}^{(s)}}{v_+^{(s)}}
\big|t_R^{s,\rightarrow}\big|^2 + \frac{v_{\text{ext},-}^{(s)}}{v_-^{(s)}} \big|t_L^{s,\leftarrow}\big|^2 \right).
\end{equation}
The total width combines resonant round‑trip dynamics with tunneling probabilities weighted by flux‑matching velocity factors. We separate it into two outgoing partial widths:
\begin{align}
\Gamma_{R,\text{out}}^{(s)} &= \frac{\hbar}{\tau_{\text{round}}^{(s)}}\frac{v_{\text{ext},+}^{(s)}}{v_+^{(s)}}
\big|t_R^{s,\rightarrow}\big|^2, \\
\Gamma_{L,\text{out}}^{(s)} &= \frac{\hbar}{\tau_{\text{round}}^{(s)}}\frac{v_{\text{ext},-}^{(s)}}{v_-^{(s)}} \big|t_L^{s,\leftarrow}\big|^2,
\end{align}
so that
\begin{equation}
\Gamma_\text{tot}^{(s)} = \Gamma_{L,\text{out}}^{(s)} + \Gamma_{R,\text{out}}^{(s)}.
\end{equation}

Incoming tunneling couplings describing carrier injection from reservoirs follow from time‑reversal of outgoing tunneling processes. Reversing propagation directions and transforming spin degrees of freedom yields
\begin{align}
\Gamma_{R,\text{in}}^{(s)} &= \frac{\hbar}{\tau_{\text{round}}^{(s)}}\frac{v_-^{(s)}}{v_{\text{ext},-}^{(s)}}
\big|t_R^{s,\leftarrow}\big|^2, \\
\Gamma_{L,\text{in}}^{(s)} &= \frac{\hbar}{\tau_{\text{round}}^{(s)}}\frac{v_+^{(s)}}{v_{\text{ext},+}^{(s)}} \big|t_L^{s,\rightarrow}\big|^2.
\end{align}
These incoming couplings do not contribute to quasi‑bound‑state decay; instead, they set injection rates entering the master‑equation description of steady‑state transport.

To summarize the physical roles of the fields: the Zeeman field $B$ both splits resonant spin levels and modifies transmission 
amplitudes and hence tunneling widths. Rashba coupling $\alpha_R$, by contrast, produces no linear shift of resonance energy. 
Its influence manifests through wavevectors, group velocities, and tunneling matrix elements. Nonzero quadratic nonreciprocal 
conductance $G_2$ requires both finite $\alpha_R$ and finite $B$.

In the general case with nonzero $B$ and $\alpha_R$, the four couplings $\Gamma_{L,\text{in}}^{(s)},\Gamma_{R,\text{in}}^{(s)},\Gamma_{L,\text{out}}^{(s)},\Gamma_{R,\text{out}}^{(s)}$ are independent parameters. At $B=0$, 
the combined symmetry of time‑reversal and geometric mirror operations imposes
\[
\Gamma_{L,\text{in}}^{(s)}=\Gamma_{R,\text{in}}^{(-s)},\qquad
\Gamma_{L,\text{out}}^{(s)}=\Gamma_{R,\text{out}}^{(-s)}.
\]
Finite $B$ breaks time‑reversal symmetry, lifting these equalities and enabling nonreciprocal charge transport. If Rashba 
coupling is also switched off ($\alpha_R\to0$), wavevectors and group velocities become symmetric, and the four couplings for 
each spin branch collapse, recovering conventional resonant‑level model behavior.

\section{Discussion on nonreciprocal conductance in the noninteracting limit}

\begin{figure}[htbp]
\centering
\includegraphics[width=0.6\columnwidth]{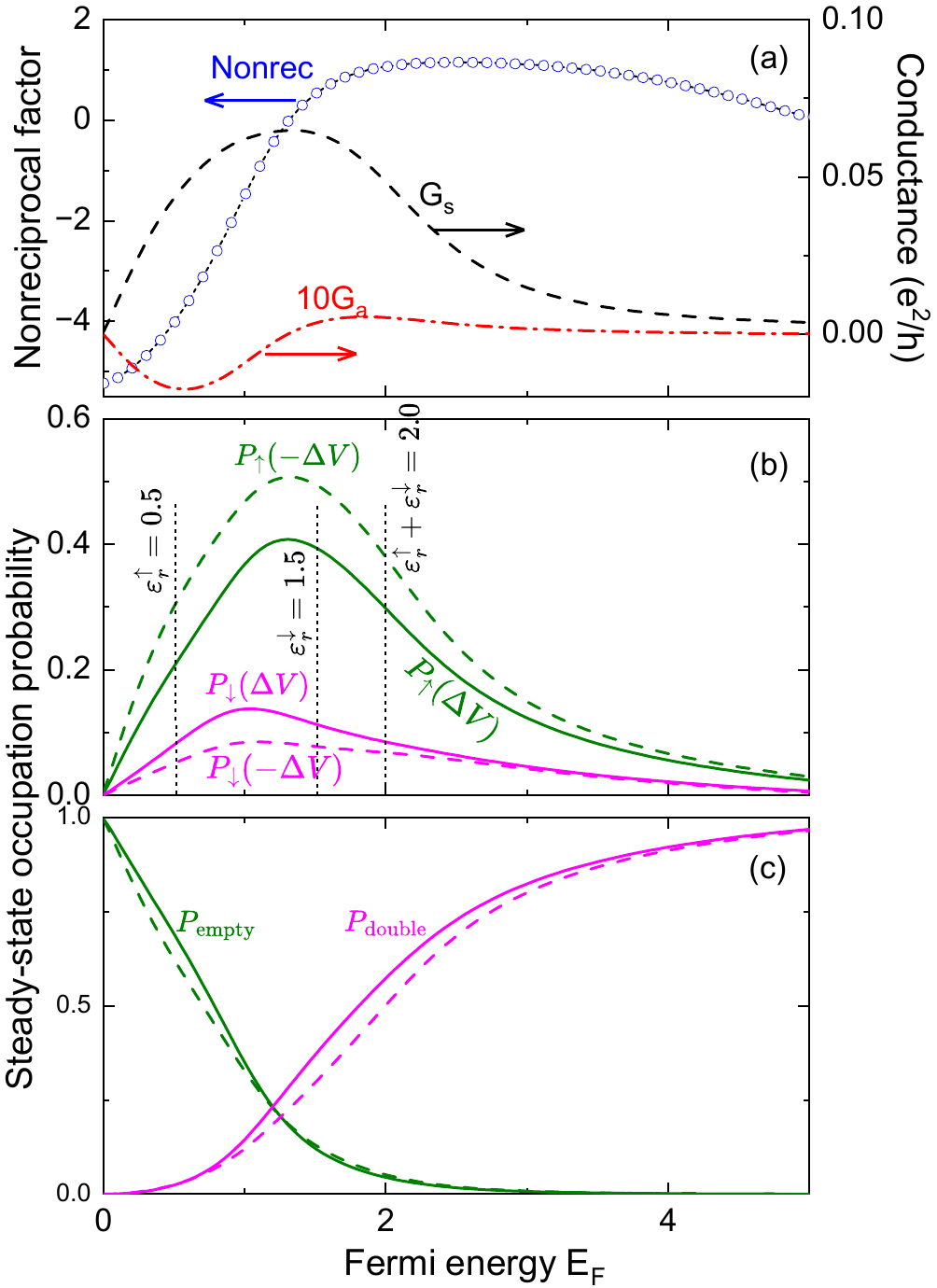}
\caption{(Color online) (a) Nonreciprocal factor, symmetric conductance $G_s$, and antisymmetric conductance $10G_a$. (b) Steady‑state occupation probabilities $P_\uparrow(\pm\Delta V)$ and $P_\downarrow(\pm\Delta V)$ for forward and reversed bias. (c) Steady‑state probabilities for empty $P_{\rm empty}$ and doubly occupied $P_{\rm double}$ states. Insets illustrate bias‑dependent occupations originating from asymmetric left‑right tunneling rates. Parameters are identical to those used in Fig.\ 4 of the main text.}
\label{fig:S1}
\end{figure}

Figure~\ref{fig:S1} shows numerical results obtained with the same 
parameters as Fig.~4 but with the Coulomb interaction set to \(U=0\).
As shown in panel (a), the nonreciprocal factor takes moderate negative 
values of roughly \(-5\%\) only within a narrow low-\(E_F\) window and remains 
small over most of the scanned Fermi-energy range. As \(E_F\) sweeps across the two resonant energies, the nonreciprocal 
factor changes sign but stays weak in magnitude. By contrast, the interacting model 
also exhibits sign reversal yet yields much larger nonreciprocal amplitudes.

At low \(E_F\), the double-occupation probability is suppressed, so Coulomb interactions barely modify transport properties. 
Accordingly, interacting and noninteracting results agree with each other 
in this low-energy regime below resonance. We do not extend 
calculations to very large \(E_F\): both symmetric and antisymmetric conductances decay, rendering the nonreciprocal factor ill-defined.

Panels (b) and (c) present occupation probabilities of the many-body configurations: \(P_\uparrow\) for the \(|10\rangle\) 
state, \(P_\downarrow\) for the \(|01\rangle\) state, \(P_\text{empty}\) for the empty \(|00\rangle\) 
configuration, and \(P_\text{double}\) for the doubly occupied \(|11\rangle\) configuration. 
Here \(\pm\Delta V\) denotes forward and reverse bias.

Compared with the interacting case in the main text (Fig.~4), bias reversal $\Delta V\to -\Delta V$
in the noninteracting calculation induces a substantial variation in \(P_\uparrow\), while \(P_\downarrow\) 
changes only modestly, so their variations no longer balance one another [Fig.~\ref{fig:S1}(b)]. 
Meanwhile, \(P_{\rm empty}\) and \(P_{\rm double}\) undergo a notable redistribution of probability weight.
Probability diverted into empty or doubly occupied configurations cannot sustain large nonreciprocal signals. 
This behavior contrasts sharply with the Coulomb-blocked interacting regime, where probability oscillates 
between \(P_\uparrow\) and \(P_\downarrow\) in a seesaw manner and generates strong nonreciprocity.

Furthermore, double occupation is allowed for \(U=0\), so \(G_s\) develops a pronounced 
peak near \(E_F\approx2\)–3. In the interacting Coulomb-blocked case, 
by contrast, a valley appears in this same energy window. Collectively, these 
features explain why nonreciprocal effects remain weak in the absence of 
Coulomb interactions. At relatively high Fermi energies, the signal magnitude 
is roughly one order of magnitude smaller than that found in the Coulomb-blockade regime
for the interacting case.

It follows that nonreciprocal transport can emerge within the master-equation 
framework once time-reversal and mirror symmetries are broken, even without 
electron-electron interactions. This mechanism is absent from the Landauer–Büttiker formalism.
The master equation accommodates internal configurational degrees of freedom 
coupled to thermal reservoirs, with relaxation and diffusive-like dynamics that 
allow nontrivial population correlations to build up in the scattering region. 
In contrast, the standard Landauer–Büttiker picture describes fully coherent elastic 
scattering: the scattering region and electrodes form a single unified wavefunction 
channel with no internal population degrees of freedom or in-situ relaxation, and 
thus cannot support the correlation-induced nonreciprocal transport reported here.

Fundamentally, the master equation shares a conceptual similarity with the 
Boltzmann transport equation. Both incorporate relaxation of distribution functions, 
reservoir coupling and dissipative transport, which naturally permit nonreciprocal 
responses under symmetry breaking. These formalisms therefore describe distinct 
transport regimes and apply to fundamentally different scattering environments.